\documentclass[
twocolumn,secnumarabic
,amssymb, amsmath,nobibnotes, aps, prl]{revtex4}
\usepackage[pdftex]{graphicx}
\usepackage{tabularx}
\usepackage{bm}%
\expandafter\ifx\csname package@font\endcsname\relax\else
 \expandafter\expandafter
 \expandafter\usepackage
 \expandafter\expandafter
 \expandafter{\csname package@font\endcsname}%
\fi

\begin{document}

\title{The ${\rm AdS}^2_{\theta}/{\rm CFT}_1$ Correspondence and Noncommutative Geometry III:\\
Phase Structure of the Noncommutative ${\rm AdS}^2_{\theta}\times \mathbb{S}^2_N$}

\author{Badis Ydri, Loubna Bouraiou}
\affiliation{Department of Physics, Badji-Mokhtar Annaba University,\\
 Annaba, Algeria.}

\begin{abstract}
The near-horizon noncommutative geometry of black holes, given by ${\rm AdS}^2_{\theta}\times \mathbb{S}^2_N$, is discussed and the phase structure of the corresponding Yang-Mills matrix models is presented. The dominant phase transition as the system cools down, i.e.  as the gauge coupling constant is decreased is an emergent geometry transition between a geometric noncommutative ${\rm AdS}^2_{\theta}\times\mathbb{S}^2_N$ phase (discrete spectrum) and a Yang-Mills matrix phase (continuous spectrum) with no background geometrical structure. We also find a possibility for topology change transitions in which space or time directions grow or decay as the temperature is varied. Indeed,  the noncommutative near-horizon geometry ${\rm AdS}^2_{\theta}\times\mathbb{S}^2_N$ can evaporate only partially to a fuzzy sphere $\mathbb{S}^2_N$ (emergence of time) or to a noncommutative anti-de Sitter spacetime ${\rm AdS}^2_{\theta}$ (topology change).

\end{abstract}

\maketitle
\tableofcontents
\section{Introduction}

The single most important fact about ${\rm AdS}^2$ geometry  is its appearance as a near-horizon geometry  of extremal black holes in both general relativity and string theory.  The typical example is Einstein gravity coupled to Maxwell electromagnetism and its celebrated four-dimensional Reissner-Nordstrom black hole given by the metric \cite{RN}

\begin{eqnarray}
&&ds^2=-f(r)d\tau^2+\frac{dr^2}{f(r)}+r^2d\Omega_2^2\nonumber\\
&&f(r)=1-\frac{2M}{r}+\frac{Q^2}{r^2}.
\end{eqnarray}
This black hole is characterized by a mass $M$ and a charge $Q$ where $M\geq Q$ (otherwise if $M< Q$ a naked singularity appears which is forbidden by cosmic censorship \cite{Penrose:1969pc}).  In the Reissner-Nordstrom black hole solution the electric field (which is not written explicitly) plays a fundamental role by supporting the whole geometry.

The near-horizon geometry of this solution is approximately  a Rindler spacetime (recall the Schwarzschild solution) which does not solve Einstein equations. However, for extremal black holes (those with mass $M=Q$ or equivalently zero temperature $T=0$) the near-horizon geometry is anti-de Sitter spacetime ${\rm AdS}^2$ (times a sphere $\mathbb{S}^2$ because of rotational invariance) which is actually an exact solution of Einstein equations. Thus, a quantum black hole with mass $M> Q$ will evaporate until it reaches the extremal mass $M=Q$ where the temperature vanishes and the evaporation stops, i.e. the extremal quantum black hole acts as a stable ground state in the case of a charged black hole \cite{Hawking:1974sw}.

In the extremal limit $M=Q$ (or $T=0$) the inner and outer horizons  $r_-$ and $r_+$ respectively coincide $r_+=r_-=Q$ and the horizon becomes a double-zero since $f(r)=(1-Q/r)^2$. We define then

\begin{eqnarray}
r=Q(1+\frac{\lambda}{z})~,~\tau=\frac{Qt}{\lambda}.
\end{eqnarray}
The near-horizon geometry of the extremal solution is obtained by letting $\lambda\longrightarrow 0$. By substituting these definitions in the metric and taking the limit $\lambda\longrightarrow 0$ we obtain

\begin{eqnarray}
ds^2=\frac{Q^2}{z^2}(-dt^2+dz^2)+Q^2(d\theta^2+\sin^2\theta d\phi^2).\label{met}
\end{eqnarray}
This is the metric of ${\rm AdS}^2\times\mathbb{S}^2$ where the charge $Q$ appears as the radius of both factors ${\rm AdS}^2$ and $\mathbb{S}^2$ \cite{carter}.

According to the proposal put forward in \cite{ydri1} the near-horizon classical geometry of a Reissner-Nordstrom black hole is actually given by the noncommutative  ${\rm AdS}^2_{\theta}\times \mathbb{S}^2_N$ where the ${\rm CFT}_1$ theory at the boundary of the noncommutative  ${\rm AdS}^2_{\theta}$ is postulated to be given by the dAFF conformal quantum mechanics, i.e. ${\rm CFT}_1\equiv{\rm QM}$. This is then a correspondence or duality between quantum mechanics (${\rm QM}$) on the boundary and noncommutative geometry (${\rm NCG}$) in the bulk which provides a concrete model for the  ${\rm AdS}^{d+1}/{\rm CFT}_d$ correspondence \cite{Maldacena:1997re} in one dimension (see figure $1$). The main argument underlying this ${\rm QM}/{\rm NCG}$ duality consists in the following main observations:
\begin{enumerate}
\item Both noncommutative ${\rm AdS}^2_{\theta}$ and the dAFF conformal quantum mechanics enjoy the same symmetry structure given by the group $SO(1,2)$. However, dAFF conformal quantum mechanics is only quasi-conformal in the sense that there is neither an invariant vacuum state nor strictly speaking primary operators  \cite{deAlfaro:1976vlx,Chamon:2011xk}. Analogously, noncommutative  ${\rm AdS}^2_{\theta}$ is only quasi-AdS as it approaches ${\rm AdS}^2$ only at large distances (commutative limit).
\item Asymptotically ${\rm AdS}^2_{\theta}$ is an ${\rm AdS}^2$ spacetime, i.e. it has the same boundary \cite{Pinzul:2017wch}. And furthermore the algebra of quasi-primary operators on the boundary (which defines in the same time the geometry of the boundary and the dAFF quantum mechanics) is in some sense a subalgebra of the operator algebra of noncommutative  ${\rm AdS}^2_{\theta}$ \cite{ydri1}.
\item Metrically the Laplacian operator on the noncommutative ${\rm AdS}^2_{\theta}$ shares the same spectrum as the Laplacian operator on the commutative ${\rm AdS}^2$ spacetime \cite{Jurman:2013ota}. The boundary correlation functions computed using the quasi-primary operators reproduces the bulk ${\rm AdS}^2$ correlation functions  \cite{Chamon:2011xk}.
\end{enumerate}
Thus here, noncommutative geometry provides the fundamental mathematical structure for the eluding ${\rm AdS}^2/{\rm CFT}_1$ correspondence \cite{Strominger:1998yg,Spradlin:1999bn}. For other possibly related approaches see \cite{Okazaki:2015lpa,Okazaki:2017lpn} and \cite{Gupta:2019cmo,Gupta:2017lwk,Gupta:2015uga,Gupta:2013ata}.

However in general, noncommutative geometry \cite{Connes:1996gi} provides a description for classical gravity while the corresponding IKKT-type Yang-Mills matrix models  \cite{Ishibashi:1996xs,Myers:1999ps} provide a proper description for a quantum theory gravity. For a related set of old and new ideas see also fuzzy physics \cite{Ydri:2001pv}  and non-perturbative lattice-like matrix-based approaches to superstring theory \cite{Hanada:2016jok}.

In this article we will focus mainly on the phase structure of the noncommutative  ${\rm AdS}^2_{\theta}\times \mathbb{S}^2_N$ within the context of the corresponding IKKT-type Yang-Mills matrix model.

This article is organized as follows. In the second section we will write down the Yang-Mills matrix models on the fuzzy sphere $\mathbb{S}^2_N$, on the noncommutative  pseudo-sphere $\mathbb{H}^2_{\theta}$, on the noncommutative anti-de Sitter spacetime ${\rm AdS}^2_{\theta}$ and on the noncommutative  near-horizon geometry ${\rm AdS}^2_{\theta}\times \mathbb{S}^2_N$. In section $3$ we study the noncommutative classical backgrounds of the fuzzy sphere $\mathbb{S}^2_N$, the noncommutative pseudo-sphere $\mathbb{H}^2_{\theta}$ and the noncommutative ${\rm AdS}^2_{\theta}$ from the matrix model and group theory perspectives and introduce a cutoff regularization of the noncommutative pseudo-sphere $\mathbb{H}^2_{\theta}$.

In section $4$ we construct explicitly the noncommutative near-horizon geometry  ${\rm AdS}^2_{\theta}\times \mathbb{S}^2_N$ and define the commutative limit.

In section $5$ we study the gravitational sector of the IKKT-type Yang-Mills matrix models in the commutative limit at the classical level (emergent gravity).

In section $6$ and $7$ we study the gauge sector of the IKKT-type Yang-Mills matrix models in the commutative limit at the quantum level by means of the effective potentials (emergent geometry and gauge theory phase structure). In particular, an emergent geometry transition is observed in all considered Yang-Mills matrix models between a geometric noncommutative geometry phase (discrete spectrum) and a Yang-Mills matrix phase (continuous spectrum) as the temperature is varied. Indeed, as the temperature is increased the background geometry evaporates to a model of pure near-commuting matrices with a uniform distribution. In the noncommutative geometry phase we have always a noncommutative gauge theory with additional Higgs-like couplings to a normal scalar field fluctuating around the classical noncommutative matrix background (a sphere, a pseudo-sphere, anti-de Sitter spacetime and near-horizon geometry of a Reissner-Nordstrom black hole).

Section $8$ contains a conclusion.

In the appendices the relation between the near-horizon geometry ${\rm AdS}^2\times\mathbb{S}^2$ and dilaton gravity is explained and the role of the non-unitary infinite-dimensional representations of $SO(1,2)$ is discussed.

\begin{figure}[htbp]
\includegraphics[width=8.0cm,angle=0]{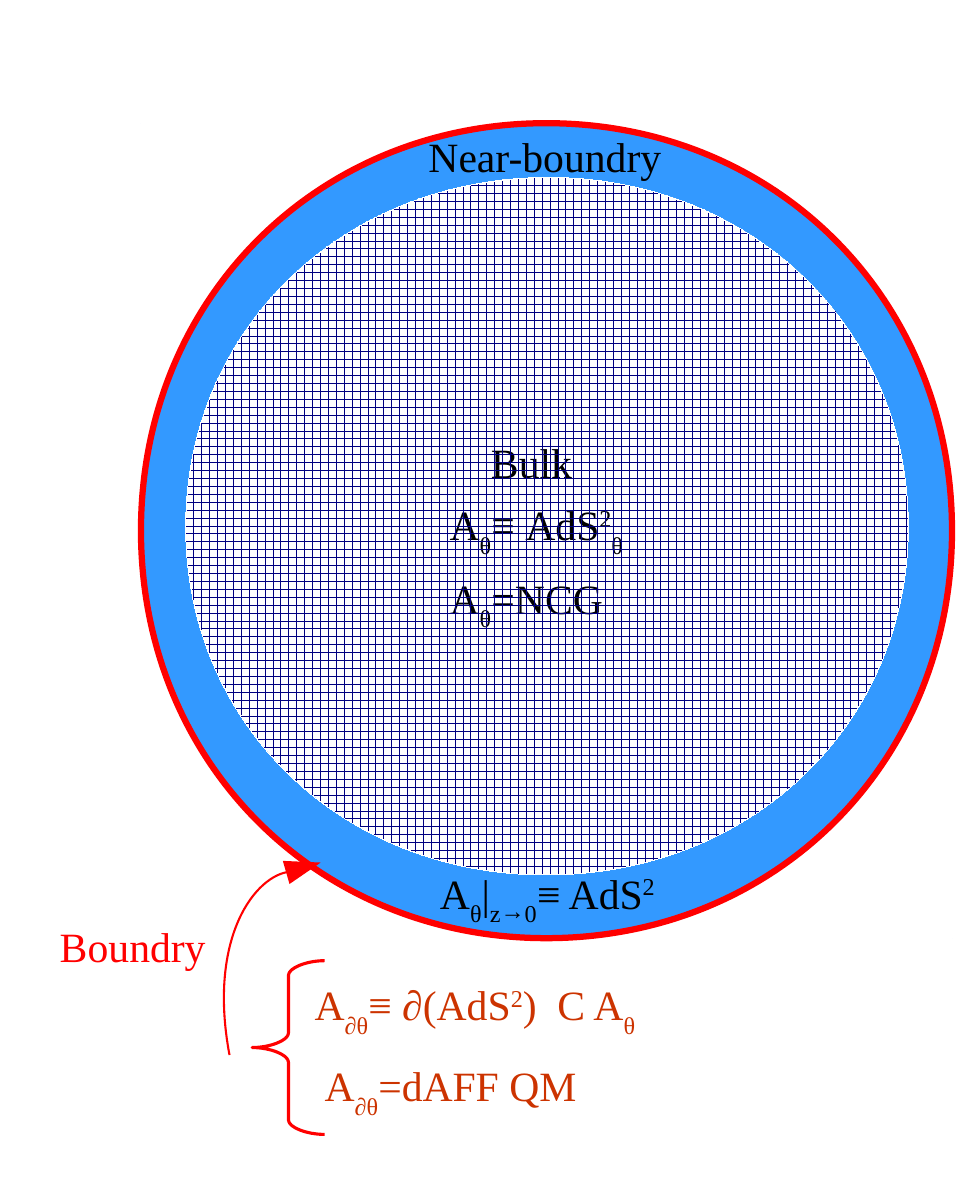}
\caption{The QM/NCG correspondence: The bulk is given by the algebra associated with noncommutative ${\rm AdS}^2_{\theta}$ while the boundary is given by a subalgebra thereof. The behavior near-boundary is precisely that of commutative ${\rm AdS}^2$. }\label{QMNCG}
\end{figure}

\section{The IKKT-type Yang-Mills matrix models}

The embedding coordinates of $\mathbb{S}^2$ and ${\rm AdS}^2$ are denoted respectively by $x^a$ and $X^a$. We will consider mostly Euclidean ${\rm AdS}^2$ which is the pseudo-sphere $\mathbb{H}^2$. These coordinates satisfy the constraints
\begin{eqnarray}
x_1^2+x_2^2+x_3^2=r^2~,~\mathbb{S}^2\in \mathbb{R}^3.
\end{eqnarray}
\begin{eqnarray}
-X_1^2+X_2^2+X_3^2=-R^2~,~\mathbb{H}^2\in \mathbb{M}^{1,2}.
\end{eqnarray}
\begin{eqnarray}
-X_1^2-X_2^2+X_3^2=-R^2~,~{\rm AdS}^2\in \mathbb{M}^{2,1}.
\end{eqnarray}
From the metric (\ref{met}) we can see that $\mathbb{S}^2$ and ${\rm AdS}^2$ are characterized by the same radius and hence we must also have
\begin{eqnarray}
R=r.
\end{eqnarray}
We can now immediately write down the Yang-Mills matrix models which enjoy the fuzzy $\mathbb{S}^2_N$, the noncommutative  ${\rm AdS}^2_{\theta}$ and the noncommutative  ${\rm AdS}^2_{\theta}\times \mathbb{S}^2_N$ respectively as their global minimum. These Yang-Mills matrix models are essentially truncation of the IKKT matrix model \cite{Ishibashi:1996xs} to lower dimensions. But they involve in an essential way a cubic Myers term \cite{Myers:1999ps} which is responsible for the condensation or emergence of matrix/noncommutative geometry. The fuzzy $\mathbb{S}^2_N$, the noncommutative  ${\rm AdS}^2_{\theta}$ and the noncommutative  ${\rm AdS}^2_{\theta}\times \mathbb{S}^2_N$ present "first quantization" of the classical (commutative) geometry of  the commutative sphere $\mathbb{S}^2$, the commutative  anti-de Sitter ${\rm AdS}^2$ and the commutative  near-horizon geometry ${\rm AdS}^2\times\mathbb{S}^2$ respectively whereas the corresponding Yang-Mills matrix models present "second quantization" which captures the quantum gravitational fluctuations around these noncommutative backgrounds.

These Yang-Mills IKKT-type matrix models are given respectively by the following three actions (the first two are $D=3$ matrix models while the third is a $D=6$ matrix model)

\begin{eqnarray}
S_{\rm S}[C]&=&N_{\rm S} {\rm Tr}_{\rm S}L_{\rm S}[C]~,~\mathbb{S}^2.\label{action1}
\end{eqnarray}
\begin{eqnarray}
S_{\rm H}[D]&=&N_{\rm H} {\rm Tr}_{\rm H}L_{\rm H}[D]~,~\mathbb{H}^2.\label{action2}
\end{eqnarray}
\begin{eqnarray}
S_{\rm HS}[D,C]&=&N_{\rm H} N_{\rm S}{\rm Tr}_{\rm H}{\rm Tr}_{\rm S}\bigg(L_{\rm S}[C]+L_{\rm H}[D]\nonumber\\
&-&\frac{1}{4}[D_a,C_b][D^a,C_b]\bigg)~,~\mathbb{H}^2\times\mathbb{S}^2.\label{action3}
\end{eqnarray}
The Lagrangian terms are given in terms of three $(2l+1)\times (2l+1)$ Hermitian matrices $C_a$ (and thus the trace ${\rm Tr}_{\rm S}$ is finite dimensional, i.e. ${\rm Tr}_{\rm S}=2l+1\equiv N_{\rm S}$) and three Hermitian operators $D_a$ (and thus the trace ${\rm Tr}_{\rm H}$ is infinite dimensional)  by the equations
\begin{eqnarray}
L_{\rm S}[C]&=&-\frac{1}{4}[C_a,C_b]^2+\frac{2i}{3}\alpha \epsilon_{abc}C_aC_bC_c\nonumber\\
&+&\beta_S C_a^2.
\end{eqnarray}
\begin{eqnarray}
L_{\rm H}[D]&=&-\frac{1}{4}[D_a,D_b][D^a,D^b]+\frac{2i}{3}\kappa f_{abc}D^aD^bD^c\nonumber\\
&+&\beta_H D_aD^a.
\end{eqnarray}
The action $S_{\rm S}[C]$ in the sphere sector is considered for example in \cite{Alekseev:2000fd,Iso:2001mg,Steinacker:2003sd}.  The ambient metric in the sphere sector is naturally Euclidean and the Levi-Civita tensor $\epsilon_{abc}$ provides the structure constants of the rotation group $SO(3)=SU(2)/\mathbb{Z}_2$. 

The ambient metric in the ${\rm AdS}$ sector is Lorentzian given by $\eta=(-1,+1,+1)$ (for Euclidean ${\rm AdS}^2$, i.e. for the pseudo-sphere $\mathbb{H}^2$) or by $\eta=(-1,-1,+1)$ (for Lorentzian ${\rm AdS}^2$) and as a consequence $f_{abc}$ are the structure constants of $SO(1,2)=SU(1,1)/\mathbb{Z}_2$ or $SO(2,1)=SU(1,1)/\mathbb{Z}_2$ respectively with a Lie algebra given by $su(1,1)$ in both cases.


The classical equations of motion which follow from the sphere action $S_{\rm S}[C]$ and the pseudo-sphere action $S_{\rm H}[D]$ are given respectively by 
\begin{eqnarray}
[C^b,B_{ab}]-2\beta_S C_a=0~,~B_{ab}=[C_a,C_b]-i\alpha \epsilon_{abc}C^c~,~\mathbb{S}^2.\nonumber\\
\end{eqnarray}
\begin{eqnarray}
[D^b,F_{ab}]-2\beta_H D_a=0~,~F_{ab}=[D_a,D_b]-i\kappa f_{abc}D^c~,~\mathbb{H}^2.\label{adsS}\nonumber\\
\end{eqnarray}
A solution of these equations of motion is given by
\begin{eqnarray}
C_a=\phi_S L_a~,~\phi_S=\alpha\varphi_S.\label{S}
\end{eqnarray}
\begin{eqnarray}
D_a=\phi_H K_a~,~\phi_H=\kappa\varphi_H.\label{H}
\end{eqnarray}
The order parameters $\varphi_S$ and $\varphi_H$ are functions of the parameters $\tau_S=\beta_S/\alpha^2$ and  $\tau_H= -\beta_H/\kappa^2$ respectively. Explicitly, we have
\begin{eqnarray}
\varphi^3-\varphi^2+\tau\varphi=0\Rightarrow \varphi_0=0~,~\varphi_{\pm}=\frac{1}{2}\big(1\pm\sqrt{1-4\tau}\big).\nonumber\\
\end{eqnarray}
The matrices $L_a$ and the operators $K_a$ appearing in (\ref{S}) and (\ref{H}) are essentially what defines the fuzzy sphere $\mathbb{S}^2_N$ and the noncommutative pseudo-sphere $\mathbb{H}^2_{\theta}$ (and the noncommutative anti-de Sitter spacetime ${\rm AdS}^2_{\theta}$) respectively.
\section{The noncommutative ${\rm AdS}^2_{\theta}$ and the fuzzy sphere $\mathbb{S}^2_N$}
The  matrices $L_a$ appearing in the solution (\ref{S}) are the angular momentum operators which generates the group of rotations $SO(3)$ of the sphere $\mathbb{S}^2$. They satisfy the $su(2)$ Lie algebra

\begin{eqnarray}
  [L_a,L_b]=i\epsilon_{abc}L_c.\label{L}
\end{eqnarray}
The irreducible representations of this algebra are characterized by the eigenvalues $l(l+1)$ of the Casimir operator
\begin{eqnarray}
  C=L_1^2+L_2^2+L_3^2.
\end{eqnarray}
It is not difficult to show that the allowed values are $l=0,1/2,1,3/2,2....$ with degeneracy equals $N=2l+1$ for each spin $l$ representation which can be labeled by the eigenvalues $m$ of $L_3$ given by $l,l-1,...,-l+1,-l$. In other words, the corresponding Hilbert spaces is given by ${\cal H}_l=\{|lm\rangle\}$.

Similarly, the  operators $K_a$ appearing in the solution (\ref{H}) are the generators of the group of pseudo-rotations $SO(1,2)$ of the pseudo-sphere $\mathbb{H}^2$. They satisfy the $su(1,1)$ Lie algebra

\begin{eqnarray}
  [K_a,K_b]=if_{ab}~^cK_c.\label{K}
\end{eqnarray}
The structure constants are given by $f^{ab}~c=-\epsilon^{ab}~c$ for Lorentzian ${\rm AdS}^2$ and $f^{ab}~c=\epsilon^{ab}~c$ for Euclidean ${\rm AdS}^2$.

The irreducible representations of the above $su(1,1)$ algebra are characterized by the eigenvalues $\pm k(k-1)$ of the Casimir operator
\begin{eqnarray}
  C=-K_1^2\mp K_2^2+K_3^2.
\end{eqnarray}
The plus sign corresponds to Lorentzian  ${\rm AdS}^2$ whereas the minus sign corresponds to Euclidean  ${\rm AdS}^2$.

There are several classes of irreducible representations of $su(1,1)$ given by the following \cite{barg,bns}:
\begin{itemize}
\item The discrete series $D^{\pm}_k$ with $k=\{1/2,1,3/2,2,...\}$. The corresponding Hilbert spaces are ${\cal H}_k=\{|km\rangle, m=\pm k,\pm (k+1), \pm(k+2),...\}$.
\item The continuous series $C_k^{\frac{1}{2}}\equiv P_s^{\frac{1}{2}}$ with $k=\frac{1}{2}+is$ where $s$ is a real number. The Hilbert spaces are given in this case by $ {\cal H}_k=\{|km+\frac{1}{2}\rangle; m=0,\pm 1,\pm 2,...\}$.
\item The complementary series $C_k^{0}\equiv P_k^0$ with $k$ a real number in the range $[0,1]$ and with Hilbert spaces $ {\cal H}_k=\{|km\rangle; m=0,\pm 1,\pm 2,...\}.$.
\item The finite-dimensional non-unitary irreducible representations $F_k$ of $su(1,1)$ with $k-1\in \mathbb{N}/2$. These coincide with the irreducible representations  of $su(2)$ with a spin quantum number $j=k-1$. 
\end{itemize}
The traces  ${\rm Tr}_{\rm S}$ and  ${\rm Tr}_{\rm S}$ appearing in the actions (\ref{action1}), (\ref{action2}) and (\ref{action3}) should now be understood to be defined in the Hilbert spaces ${\cal H}_l$ and ${\cal H}_k$ respectively. In particular, the trace  ${\rm Tr}_{\rm S}$ is finite dimensional, i.e. $ {\rm Tr}_{\rm S}{\bf 1}=N$ and thus we must have  the identification
\begin{eqnarray}
N_{\rm S}\equiv N=2l+1.\label{Sreg}
\end{eqnarray}
The ${\rm AdS}$ trace ${\rm Tr}_{\rm H}$ on the other hand will be regularized  in such a way that only $2k-1$ states are included, i.e.  ${\rm Tr}_{\rm H}{\bf 1}=2k-1$ and correspondingly we will choose the overall normalization $N_{\rm H}$ such that $N_{\rm H}\equiv 2k-1$ for a complete parallel with the sphere sector  \cite{ydri2}. Thus, the pseudo-spin quantum number $k-1$ is the analogue of the spin quantum number $l$ and the operators $D_a$ become  therefore $(2k-1)\times (2k-1)$ matrices. This regularization is discussed further in  \cite{ydri2}. We have then
\begin{eqnarray}
N_{\rm H}\equiv 2k-1.\label{reg}
\end{eqnarray}
From the metric (\ref{met}) we see that  the ${\rm AdS}$ spacetime has the same radius as the sphere and hence we will also impose the natural identification 
\begin{eqnarray}
N_{\rm H}\equiv N.\label{Hreg}
\end{eqnarray}
The coordinate operators $\hat{x}_a$ on the fuzzy sphere $\mathbb{S}^2_N$ are defined by
\begin{eqnarray}
C_a=\varphi_S \hat{x}_a.\label{S1}
\end{eqnarray}
We want them to satisfy the embedding relation
\begin{eqnarray}
\hat{x}_1^2+\hat{x}_2^2+\hat{x}_3^2=r^2~,~\mathbb{S}^2_N.\label{emb1}
\end{eqnarray}
This holds if and only if
\begin{eqnarray}
\frac{r^2}{\alpha^2}=l(l+1).
\end{eqnarray}
They also satisfy
\begin{eqnarray}
[\hat{x}_a,\hat{x}_b]=i\alpha\epsilon_{abc}\hat{x}_c.\label{comm1}
\end{eqnarray}
Similarly, the coordinate operators $\hat{X}_a$ on the noncommutative pseudo-sphere $\mathbb{H}^2_{\theta}$ and on the noncommutative ${\rm AdS}^2_{\theta}$ are defined by
\begin{eqnarray}
D_a=\varphi_H \hat{X}_a.\label{H1}
\end{eqnarray}
We want them to satisfy the embedding relations
\begin{eqnarray}
&&-\hat{X}_1^2+\hat{X}_2^2+\hat{X}_3^2=-R^2~,~\mathbb{H}^2_{\theta}\nonumber\\
&&-\hat{X}_1^2-\hat{X}_2^2+\hat{X}_3^2=-R^2~,~{\rm AdS}^2_{\theta}.\label{emb2}
\end{eqnarray}
This is indeed true if and only if
\begin{eqnarray}
\frac{R^2}{\kappa^2}=\pm k(k-1).
\end{eqnarray}
The minus sign corresponds to Lorentzian  ${\rm AdS}^2$ whereas the plus sign corresponds to Euclidean  ${\rm AdS}^2$ or $\mathbb{H}^2$. These coordinate operators also satisfy
\begin{eqnarray}
[\hat{X}_a,\hat{X}_b]=i\kappa f_{abc}\hat{X}^c.\label{comm2}
\end{eqnarray}
In each case the solution of both the commutator equation and the embedding condition are given by the irreducible representations of the corresponding Lie algebra.

Explicitly, we have:
\begin{enumerate}
  \item Fuzzy sphere $\mathbb{S}^2_N$: In this case the solution is given by \cite{Hoppe,Madore:1991bw}
  \begin{eqnarray}
    \hat{x}_a=\alpha L_a.
  \end{eqnarray}
\item Noncommutative Lorentzian ${\rm AdS}^2_{\theta}$: In this case the solution is given by \cite{Ho:2000fy,Ho:2000br,Jurman:2013ota}
  \begin{eqnarray}
    \hat{X}^a=\kappa K^a.
  \end{eqnarray}
  $K^a$ are the generators of the Lie algebra $su(1,1)$ which must be i) unitary (thus the finite-dimensional representations are excluded) ii) have negative Casimir by the embedding condition (and therefore the discrete series are excluded) and iii) must admit a commutative limit (and hence the complementary series are excluded since their Casimir is in a finite range). Therefore, $K^a$ are the generators of $su(1,1)$ in the continuous series $C_k^{\frac{1}{2}}=P_s^{\frac{1}{2}}$.

  \item Noncommutative Euclidean ${\rm AdS}^2_{\theta}$ or noncommutative pseudo-sphere $\mathbb{H}^2_{\theta}$: In this case the solution is given by \cite{Pinzul:2017wch}
  \begin{eqnarray}
    \hat{X}^a=\kappa K^a.
  \end{eqnarray}
  $K^a$ are the generators of the Lie algebra $su(1,1)$ which must be i) unitary (again the finite-dimensional representations are excluded) ii) have negative Casimir by the embedding condition (and therefore it is the continuous and complementary series which are excluded now) and iii) must admit a commutative limit (the complementary series is also excluded on this account). Therefore, $K^a$ are the generators of $su(1,1)$ in the discrete series $D_k^{\pm}$.
  \end{enumerate}

\section{The noncommutative ${\rm AdS}^2_{\theta}\times \mathbb{S}^2_N$}
We are now in a position to write down more explicitly the coordinate operators on the near-horizon noncommutative geometry  ${\rm AdS}^2_{\theta}\times \mathbb{S}^2_N$ as the tensor product of the operator algebras and Hilbert spaces associated with  the noncommutative anti-de Sitter space ${\rm AdS}^2_{\theta}$ and the fuzzy sphere $\mathbb{S}^2_N$.

First, we must remember that the sphere regulator  (\ref{Sreg}) is a fully $SO(3)-$symmetric regulator as opposed to the pseudo-sphere regulator  (\ref{Hreg}) which is an ordinary hard cutoff which breaks $SO(1,2)$ invariance. These two cutoffs are given explicitly by $N_{\rm S}=2l+1$ and $N_{\rm H}=2k-1$. Thus, the size $N$ of the matrix model (\ref{action3}) is given by
\begin{eqnarray}
N=N_{\rm S}N_{\rm H}=(2l+1)(2k-1).\label{N}
\end{eqnarray}
We can even employ the choice (\ref{Hreg}) in which case we will have $N_{\rm H}=N_{\rm S}$ or equivalently $k=l-1$ and $N=N_{\rm S}^2=(2l+1)^2$.

The solution of the equations of motion which follow from the action $S_{\rm HS}[D,C]$, which is given by equation (\ref{action3}), is given immediately by  
\begin{eqnarray}
C_a=\phi_S L_a\otimes {\bf 1}_H~,~D_a={\bf 1}_S\times\phi_H K_a.\label{HS}
\end{eqnarray}
These are $N_{\rm S}N_{\rm H}\times N_{\rm S}N_{\rm H}$ Hermitian matrices. Since the action (\ref{action3}) is well behaved only in the Euclidean signature we are only going to consider here the case of the noncommutative Euclidean ${\rm AdS}^2_{\theta}$, i.e. the case of the noncommutative pseudo-sphere $\mathbb{H}^2_{\theta}$. In other words, the $K_a$  in the above equation are the generators of $su(1,1)$ in the discrete series $D_k^{\pm}$.

The coordinate operators $(\hat{x}^a$,$\hat{X}^a)$ on the noncommutative  ${\rm AdS}^2_{\theta}\times\mathbb{S}^2_N$ are then given explicitly by 
\begin{eqnarray}
C_a=\varphi_S \hat{x}_a~,~D_a=\varphi_H \hat{X}_a.
\end{eqnarray}
Equivalently
 \begin{eqnarray}
    \hat{x}_a=\alpha L_a\otimes {\bf 1}_H~,~\hat{X}^a={\bf 1}_S\otimes \kappa K^a.
  \end{eqnarray}
  These coordinate operators satisfy the embedding relations and commutator equations given by 

\begin{eqnarray}
&&-\hat{X}_1^2+\hat{X}_2^2+\hat{X}_3^2=-R^2~,~\hat{x}_1^2+\hat{x}_2^2+\hat{x}_3^2=r^2.
\end{eqnarray}
\begin{eqnarray}
&&[\hat{X}_a,\hat{X}_b]=i\kappa f_{abc}\hat{X}^c\nonumber\\
&&[\hat{x}_a,\hat{x}_b]=i\alpha\epsilon_{abc}\hat{x}_c\nonumber\\
&&[\hat{x}_a,\hat{X}_b]=0.
\end{eqnarray}
Again, we must also have the quantization of the two deformation parameters $\alpha$ and $\kappa$ in terms of the radii $R$ and $r$ and in terms of the spin and pseudo-spin quantum numbers $l$ and $k$, viz 

\begin{eqnarray}
\frac{r^2}{\alpha^2}=l(l+1)~,~\frac{R^2}{\kappa^2}=k(k-1).
\end{eqnarray}
By expanding the sphere action (\ref{action1}) around the background (\ref{S}) we obtain a $U(1)$ gauge field on the fuzzy sphere $\mathbb{S}^2_N$ where the gauge coupling constant $1/g^2_S$ is given by
\begin{eqnarray}
\frac{1}{g^2_S}=\tilde{\alpha}^4=4\alpha^4l(l+1).\label{Sg}
\end{eqnarray}
The full gauge dynamics obtained is actually a noncommutative $U(1)$ gauge field on the fuzzy sphere coupled to a scalar field normal to the sphere, i.e. a noncommutative Higgs theory on $\mathbb{S}^2_N$. See for example \cite{CastroVillarreal:2004vh}. 

Similarly, by expanding the pseudo-sphere action (\ref{action2}) around the background (\ref{H}) we obtain a $U(1)$ gauge field on the noncommutative pseudo-sphere $\mathbb{H}^2_N$ where the gauge coupling constant $1/g^2_H$ is given by
\begin{eqnarray}
\frac{1}{g^2_H}=\tilde{\kappa}^4=4\kappa^4k(k-1).\label{Hg}
\end{eqnarray}
Again, the full gauge dynamics is a noncommutative $U(1)$ gauge field on the noncommutative pseudo-sphere coupled to a scalar field normal to the pseudo-sphere, i.e. a noncommutative Higgs theory on Euclidean ${\rm AdS}^2_{\theta}$.

Next we expand the action (\ref{action3}) around the background (\ref{HS}) to obtain a four-dimensional noncommutative gauge theory on the noncommutative near-horizon geometry ${\rm AdS}^2_{\theta}\times\mathbb{S}^2_N$ with a gauge coupling constant given in the sphere and pseudo-sphere sectors by \cite{Ydri:2016osu} 
\begin{eqnarray}
\frac{1}{g^2_{\rm HS}}&=&\bar{\kappa}^4=\kappa^4 k(k-1)~,~\frac{1}{g^2_{\rm HS}}=\bar{\alpha}^4=\alpha^4 l(l+1).\nonumber\\
\end{eqnarray}
Hence, in order to get a uniform gauge coupling constant across the sphere and the pseudo-sphere sectors we must have the constraint (we have also used here the choice (\ref{Hreg}))
\begin{eqnarray}
\frac{1}{g^2_{\rm HS}}=\bar{\alpha}^4=\bar{\kappa}^4=\frac{\tilde{\kappa}^4}{4N}=\frac{\tilde{\alpha}^4}{4N}.\label{gHS}
\end{eqnarray}
In this equation $\tilde{\alpha}^4=\alpha^4N^2$ and $\tilde{\kappa}^4=\kappa^4N^2$ in agreement with (\ref{Sg}) and (\ref{Hg}) respectively but with $N$ defined by (\ref{N}). This result summarizes the main difference between $2$ and $4$ dimensions, namely the fact that the gauge coupling constant in $4$ dimensions scales differently with the size of the matrices than in $2$ dimensions. 

The backgrounds (\ref{S}), (\ref{H}) and (\ref{HS}) are global minima of their corresponding Yang-Mills matrix actions yielding in the gauge sector a $U(1)$ gauge group. We can also obtain $U(n)$ gauge groups on the fuzzy sphere $\mathbb{S}^2_N$,  on the noncommutative anti-de Sitter ${\rm AdS}^2_{\theta}$ and on the noncommutative near-horizon geometry ${\rm AdS}^2_{\theta}\times \mathbb{S}^2_N$ by considering other backgrounds with smaller group representations. See for example \cite{Steinacker:2003sd}.  

The commutative limit of the sphere  $\mathbb{S}^2$ is obtained by taking $l\longrightarrow\infty$ whereas the commutative limit of anti-de Sitter spacetime ${\rm AdS}^2$ is obtained by taking $k\longrightarrow \infty$. The commutative limit of the near-horizon geometry  ${\rm AdS}^2\times \mathbb{S}^2$ is given by taking together $l\longrightarrow\infty$ and  $k\longrightarrow \infty$ in an almost obvious way. 

\section{Emergent gravity from Yang-Mills matrix models: A gravitational commutative limit}

Yang-Mills matrix models play a crucial role both in noncommutative gravity and emergent geometry. In this section we will study as an example the commutative limit of the  noncommutative ${\rm AdS}^2_{\theta}$  matrix model (\ref{action2}) in the gravitational sector. The treatment is only classical but in the next two sections we will study quantum-mechanically the commutative limit  in the gauge sector of the three matrix models (\ref{action1}), (\ref{action2}) and (\ref{action3}). 

Thus, as an example we will consider noncommutative ${\rm AdS}^2_{\theta}$ which can be obtained as the classical background solution of the $D=3$  matrix model (\ref{action2}). We will set for simplicity $\beta_H=0$.
By setting the variation of this action equals to zero we obtain the equations of motion given by (\ref{adsS}).
A solution of these equations of motion is given by (\ref{H}) or equivalently by
\begin{eqnarray}
D^a=\kappa K^a\equiv \hat{X}^a.\label{ads2}
\end{eqnarray}
The $K^a$ are the generators of the Lie group $SO(1,2)$ in the irreducible representation given by the discrete series $D_k^{\pm}$ which are labeled by an integer $k$. The $\hat{X}^a$ are thus precisely the coordinate operators on noncommutative ${\rm AdS}^2_{\theta}$ and as a consequence we have

\begin{eqnarray}
D^aD_a=\hat{X}^a\hat{X}_a=\kappa^2K^aK_a=-R^2.\label{casimir}
\end{eqnarray}
In other words, the radius $R$ of  noncommutative ${\rm AdS}^2_{\theta}$, the integer $k$ labeling the discrete series $D_k^{\pm}$ and the deformation parameter or coupling constant $\kappa$ of the corresponding Yang-Mills matrix model (\ref{action2}) are related by the condition

\begin{eqnarray}
R^2=\kappa^2k(k-1).
\end{eqnarray}
Therefore, the commutative limit $\kappa\longrightarrow 0$ corresponds to the large representation limit  $k\longrightarrow \infty$. The geometric commutative limit can be thought of as the semi-classical limit.

In noncommutative gravity the fundamental degrees of freedom of the theory are given by the hermitian  matrices $D^a$ and not by the metric $g_{\mu\nu}$ which can only emerge in the semi-classical/commutative limit $\kappa\longrightarrow 0$ (or equivalently $k\longrightarrow \infty$) as outlined in \cite{Steinacker:2008ri,Steinacker:2010rh}.

Furthermore, the noncommutativity tensor $\theta^{ab}$ (or equivalently the Poisson structure  $\theta^{\mu\nu}$ of the underlying symplectic manifold ${\rm AdS}^2$) is generically a function of the matrices $D^a$ (or equivalently of the local coordinates $x^{\mu}$ on ${\rm AdS}^2$) and plays also a more fundamental role than the emergent metric $g_{\mu\nu}$. This tensor is given explicitly by 

\begin{eqnarray}
[D^a,D^b]=i\kappa\theta^{ab}(D).\label{commutator}
\end{eqnarray}
Clearly, in the classical (classical with respect to the matrix model) configurations $D^a\equiv \hat{X}^a$ we must have $\theta^{ab}(D)\equiv f^{abc}\hat{X}_c$.

The matrix coordinates $D_a=\hat{X}^a$ behave in the commutative limit as $D_a \sim X^a$ which are the embedding coordinates of ${\rm AdS}^2$. These coordinates can always be decomposed into tangential and normal coordinates on ${\rm AdS}^2$. This can be seen  by considering for example the neighborhood of the "north pole", viz $X^3\equiv\phi\simeq R$, $X^{1}\equiv x^1 << R$ and $X^2\equiv x^2<< R$ where $x^{1}$ and $x^{2}$ are local coordinates on ${\rm AdS}^2$. The commutator (\ref{commutator}) around the "north pole" becomes then $[\hat{x}^{\mu},\hat{x}^{\nu}]=i\kappa \theta^{\mu\nu}$ where $\theta^{\mu\nu}=Rf^{\mu\nu 3}$.

We decompose then the matrices $D^a\equiv \hat{X}^a$ into tangential and normal components as

\begin{eqnarray}
\hat{X}^a=(\hat{x}^{\mu},\hat{\phi}).
\end{eqnarray}
From the requirement (\ref{casimir}) we can see that $\phi$ is a function of $\hat{x}^{\mu}$, $\mu=1,2$, i.e.

\begin{eqnarray}
\hat{\phi}=\hat{\phi}(\hat{x}).
\end{eqnarray}
Hence, the commutator (\ref{commutator}) becomes

\begin{eqnarray}
[\hat{x}^{\mu},\hat{x}^{\nu}]=i\kappa\theta^{\mu\nu}(\hat{x})~,~\theta^{\mu\nu}\equiv f^{\mu\nu3}\hat{\phi}(\hat{x}).
\end{eqnarray}
The quantized derivations (parallel and normal) on noncommutative ${\rm AdS}^2_{\theta}$ and their commutative counterparts  on  ${\rm AdS}^2$ are then given by

\begin{eqnarray}
\hat{e}^a(F)=-i[D^a,F]\longrightarrow e^a(F)=\kappa\theta^{\mu\nu}\partial_{\mu}x^a\partial_bF.
\end{eqnarray}
Next, we introduce a covariant scalar action on noncommutative ${\rm AdS}^2_{\theta}$ by the equation

\begin{eqnarray}
S[D,\hat{\Phi}]=\frac{2\pi R \kappa}{2}Tr\bigg(-\frac{1}{R^2\kappa^2}[D^a,\hat{\Phi}][D_a,\hat{\Phi}]+m^2\hat{\Phi}^2\bigg).\label{actioncov}\nonumber\\
 \end{eqnarray}
We can now compute in the configurations $D_a=\hat{X}^a$ the kinetic term

  \begin{eqnarray}
    -\eta_{ab}[D^{a},\hat{\Phi}][D^{b},\hat{\Phi}]&=&\eta_{ab}\hat{e}^{a}(\hat{\Phi})\hat{e}^{b}(\hat{\Phi})\nonumber\\
    &\sim &\kappa^{2}\theta^{\mu\mu^{\prime}}\theta^{\nu\nu^{\prime}}{g}_{\mu\nu}{\partial}_{\mu^{\prime}}\Phi{\partial}_{\nu^{\prime}}\Phi\nonumber\\
    &\sim &{G}^{\mu^{\prime}\nu^{\prime}}{\partial}_{\mu^{\prime}}\Phi{\partial}_{\nu^{\prime}}\Phi.
  \end{eqnarray}
The quantity $G^{\mu\nu}$ is the induced metric which couples to the matter field $\Phi$ and which is given explicitly by

   \begin{eqnarray}
        {G}^{\mu^{\prime}\nu^{\prime}} &=&\kappa^{2}\theta^{\mu\mu^{\prime}}\theta^{\nu\nu^{\prime}}{g}_{\mu\nu}.
   \end{eqnarray}
Whereas $g_{\mu\nu}$ is  the embedding metric (the metric on ${\rm AdS}^2$ viewed as a Poisson manifold) given explicitly by

  \begin{eqnarray}
    {g}_{\mu\nu}&=& \eta_{ab}{\partial}_{\mu}x^{a}{\partial}_{\nu}x^{b}.
  \end{eqnarray}
The kinetic action is then given by

\begin{eqnarray}
    -Tr [D_{a},\hat{\Phi}][D^{b},\hat{\Phi}]
    &\sim &\frac{1}{2\pi}\int d^2x\rho(x){G}^{\mu^{\prime}\nu^{\prime}}{\partial}_{\mu^{\prime}}\Phi{\partial}_{\nu^{\prime}}\Phi.\label{kaction}\nonumber\\
  \end{eqnarray}
We introduced in this last equation a scalar density $\rho$, which  defines on the quantized  Poisson manifold ${\rm AdS}^2_{\theta}$ a local non-commutativity scale, by the relation

\begin{eqnarray}
 \rho=\frac{1}{\sqrt{{\rm det}{\kappa \theta^{\mu\nu}}}}.\label{sd1}
  \end{eqnarray}
The kinetic action (\ref{kaction}) does not have the canonical covariant form which can be reinstated by a rescaling of the metric as follows

\begin{eqnarray}
  \tilde{G}^{ab}=\exp(-\sigma)G^{ab}.\label{sd2}
\end{eqnarray}
And imposing the condition

 \begin{eqnarray}
\rho G^{ab}=\sqrt{{\rm det}\tilde{G}_{ab}}\tilde{G}^{ab}\Rightarrow \rho=\sqrt{{\rm det}G_{ab}}.\label{sd3}
\end{eqnarray}
By using equations  (\ref{sd1}) and (\ref{sd3}) we can show that the scalar density $\rho$ can also be written in the form $\rho=\sqrt{{\rm det}g_{ab}}$. Hence, we must have

\begin{eqnarray}
G_{ab}\equiv g_{ab}.
\end{eqnarray}
And by substituting in (\ref{sd2}) we obtain

\begin{eqnarray}
\tilde{G}^{ab}\equiv e^{-\sigma} g^{ab}.
\end{eqnarray}
We get immediately in the semi-classical limit $\kappa\longrightarrow 0$ the kinetic action

\begin{eqnarray}
    -Tr [D_{a},\hat{\Phi}][D^{b},\hat{\Phi}]
    &\sim &\frac{1}{2\pi}\int d^2x\sqrt{{\rm det}G_{\mu\nu}}{G}^{\mu^{\prime}\nu^{\prime}}{\partial}_{\mu^{\prime}}\Phi{\partial}_{\nu^{\prime}}\Phi\nonumber\\
&\sim & \frac{1}{2\pi}\int d^2x\sqrt{{\rm det}\tilde{G}_{\mu\nu}}{\tilde{G}}^{\mu^{\prime}\nu^{\prime}}{\partial}_{\mu^{\prime}}\Phi{\partial}_{\nu^{\prime}}\Phi.\nonumber\\
  \end{eqnarray}
The conformal factor $e^{-\sigma}$ remains therefore undetermined since in two dimensions Weyl transformations of the metric $G^{\mu\nu}\longrightarrow e^{-\alpha}G^{\mu\nu}$ are  in fact symmetries of the action \cite{Jurman:2013ota}.

By going through the same steps we can now show that the Yang-Mills term (quartic term) of the matrix model (\ref{action2}) reduces, in the semi-classical/commutative limit $\kappa\longrightarrow 0$ (or equivalently $k\longrightarrow \infty$), not to the Einstein equations but to the cosmological term \cite{Jurman:2013ota}. A matrix form of the Einstein equations can also be written down but this is not necessary within the formalism of noncommutative gravity since the condensation of the geometry of ${\rm AdS}^2_{\theta}$ is in fact driven by the Myers-Chern-Simons term (cubic term)  of  (\ref{action2}) \cite{Myers:1999ps}.

Indeed, the ${\rm AdS}^2_{\theta}$ solution (\ref{ads2}) of the equation of motion (\ref{adsS}) is not unique and this solution can be made more stable by adding a potential term to the action (\ref{action2}) which implements explicitly the constraint (\ref{casimir}) such as the term \cite{CastroVillarreal:2004vh}

\begin{eqnarray}
V[D]=M^2Tr(D^aD_a+R^2)^2. \label{potential}
\end{eqnarray}
The action (\ref{action2})+(\ref{potential}) will then admit for large and positive values of $M^2$ a unique solution given by the ${\rm AdS}^2_{\theta}$ background (\ref{ads2}) which satisfies the constraint (\ref{casimir}) by construction. The expansion of the scalar action (\ref{actioncov})  around the AdS solution becomes more reliable since this background in the limit $M^2\longrightarrow \infty$ is  completely stable. Therefore, the action (\ref{action2})+(\ref{potential}) acts effectively within noncommutative gravity as an Einstein-Hilbert action.

\section{Effective potentials and phase structure of $\mathbb{S}^2_N$ and ${\rm AdS}^2_{\theta}$}
Starting now from the actions (\ref{action1}) and (\ref{action2}) we compute the one-loop effective actions on the fuzzy sphere $\mathbb{S}^2_N$ and on the noncommutative fuzzy pseudo-sphere  $\mathbb{H}^2_{\theta}$ using the background field method. The one-loop effective potential around the sphere background (\ref{S}) is computed in  \cite{CastroVillarreal:2004vh} whereas the one-loop effective potential around the pseudo-sphere background (\ref{H}) is computed (using the regularization (\ref{reg})) in  \cite{ydri2}. We obtained the effective potentials
\begin{eqnarray}
\frac{2V_S}{N_{\rm S}^2}
&=&4\alpha^4 l(l+1)\bigg[\frac{1}{4}\varphi_S^4-\frac{1}{3}\varphi_S^3+\frac{1}{2}\tau_S\varphi_S^2\bigg]+\log\varphi_S^2\nonumber\\
&&\tau_S=\frac{\beta_S}{\alpha^2}.\label{effe1}
\end{eqnarray}
\begin{eqnarray}
\frac{2V_H}{N_{\rm H}^2}
&=&4\kappa^4 k(k-1)\bigg[\frac{1}{4}\varphi_H^4-\frac{1}{3}\varphi_H^3+\frac{1}{2}\tau_H\varphi_H^2\bigg]+\log\varphi_H^2\nonumber\\
&&\tau_H=-\frac{\beta_H}{\kappa^2}.\label{effe2}
\end{eqnarray}
These potentials are of the same mathematical form and thus the discussion of the corresponding phase diagrams is effectively the same. Here, the scalar fields $\varphi_S$ and $\varphi_H$ play the role of the order parameters characterizing the phase diagrams while the role of the temperatures $T_S$ and $T_H$ is played by the gauge coupling constants squared, i.e. $T_S\equiv 1/\tilde{\alpha}^4=g_S^2$ and  $T_H\equiv 1/\tilde{\kappa}^4=g_H^2$ with  $\tilde{\alpha}^4=4\alpha^4 l(l+1)$ and  $\tilde{\kappa}^4=4\kappa^4 k(k-1)$.

Again we stress the fact that the cutoff $N_{\rm S}=N$ on the fuzzy sphere is a natural $SO(3)-$invariant cutoff whereas the cutoff $N_{\rm H}=N$  on the noncommutative pseudo-sphere is only a regulator which breaks explicitly $SO(1,2)$ invariance as the noncommutative pseudo-sphere and the noncommutative anti-de Sitter spacetime are really infinite-dimensional operator algebras. See equations (\ref{reg}) and (\ref{Hreg}) and also the discussion in \cite{ydri2}. This might be related to the fact that the matrix model (\ref{action1}) is truly an Euclidean action where all its phases are accessible by the Monte Carlo method whereas the matrix model (\ref{action2}) is only Euclidean in the sense that it gives an anti-de Sitter spacetime in the noncommutative pseudo-sphere phase.

The model on the fuzzy sphere is extensively studied by analytical and Monte Carlo methods for both $\tau_S=$ and $\tau_S\neq 0$ in \cite{Azuma:2004zq,Delgadillo-Blando:2007mqd,Delgadillo-Blando:2008cuz, Azuma:2005bj,Delgadillo-Blando:2012skh}. The phase structure in this case can be summarized as follows:

\begin{itemize}
\item We start by setting the logarithmic quantum correction to zero. The classical equation of motion admits three solutions:
\begin{eqnarray}
\varphi_0=0~,~\varphi_{\pm}=\frac{1\pm\sqrt{1-4\tau_S}}{2}.
\end{eqnarray}
The solution $\varphi_0=0$ (the Yang-Mills or matrix phase) is the global minimum (ground state) of the system in the regime $\tau_S>1/4$. The solution $\varphi_-$  (the geometric or fuzzy sphere phase) is the global minimum in the regime $0<\tau_S<1/4$. The model has no ground state for $\tau_S<0$, i.e. $\beta_S<0$. The two global minima $C_a=0$ and $C_a=\phi_-L_a$ are separated by a potential barrier whose maximum height is reached at the local maximum $\varphi_+$.

\item We should also mention here that the configuration $C_a=\phi_S J_a$ is also a local minimum of the system. The $J_a$ are the generators of $SO(3)$ in a reducible representation  characterized by the spin quantum numbers $j_i< l=(N-1)/2$ satisfying $\sum_i(2j_i+1)=N$. More precisely, we find that the configuration  $C_a=\phi_-L_a$ has a negative energy and thus lower than the zero energy of the configuration $C_a=0$ only in the regime $0<\tau_S<2/9$. This negative energy is minimized when $J_a=L_a$. In this regime the fuzzy sphere is indeed stable and the expansion of the matrix model around the fuzzy background $C_a=\phi_-L_a$ gives a noncommutative gauge theory which also includes coupling to a normal scalar field, i.e. a noncommutative Higgs system. 

\item At $\tau_S=2/9$ the two configurations  $C_a=\phi_-L_a$ and  $C_a=0$ become degenerate. Thus, in the regime $2/9<\tau_S<1/4$ the fuzzy sphere becomes unstable. The coexistence curve between the geometric fuzzy sphere phase and the Yang-Mills matrix phase asymptotes therefore to the line $\tau_S=2/9$ (and not to the line $\tau_S=1/4$) where the energy functional becomes a complete square.
\item If we include the logarithmic quantum correction the potential becomes unbounded from below near $\varphi=\varphi_0=0$, i.e. the effective potential   (\ref{effe1}) is really valid only in the fuzzy sphere phase $\varphi=\varphi_-\neq 0$. But the Yang-mills phase can still be accessed by Monte Carlo simulation of the matrix model (\ref{action1}).
\item In the quantum case the minimum $\varphi_-$ (corresponding to the geometric fuzzy sphere phase) becomes a function of both $\tau_S$ and $\tilde{\alpha}^4=4\alpha^4 l(l+1)$. The critical coexistence curve exists therefore in the $(\tau_S,\tilde{\alpha})$ plane where the local minimum $\varphi_-$ disappears. The conditions determining this curve are obviously given by $V_S^{\prime}=0$ and $V_S^{\prime\prime}=0$. Explicitly, we obtain the curve $\tilde{\alpha}_*=\tilde{\alpha}_*(\tau_S)$ defined by the equations
\begin{eqnarray}
&&\frac{1}{\tilde{\alpha_*^4}}=\frac{\varphi_*^2(\varphi_*-2\tau_S)}{8}\nonumber\\
&&\varphi_*=\frac{3}{8}(1+\sqrt{1-\frac{32\tau_S}{9}}).\label{ec}
\end{eqnarray}
Thus, as we increase $\tau_S$ from $0$ to $1/4$ the critical value $\tilde{\alpha}_*$ increases from around $2$ to infinity. Thus,  the critical temperature $T_*\equiv 1/\tilde{\alpha}_*^4=g_{S*}^2$ decreases towards zero as we increase $\tau_S$ to $1/4$. In other words, the geometric fuzzy sphere phase exists in the region of low temperatures $T$ (or large $\tilde{\alpha}$) and $\tau_S<1/4$.
\item Hence, as the temperature is increased the fuzzy sphere phase evaporates to a pure matrix phase with no background geometrical structure. In this model the geometry condenses or emerges only as the system cools.
\item These predictions, which are based on the one-loop effective potential (\ref{effe1}), are confirmed by Monte Carlo simulation only for $\tau_S<2/9$. It is observed (in Monte Carlo simulation) that the coexistence curve between the geometric fuzzy sphere phase (low temperatures) and the Yang-Mills matrix phase (high temperatures) for $2/9<\tau_S<1/4$ asymptotes very rapidly to the line $\tau_S=2/9$ for $\tilde{\alpha}>\tilde{\alpha}_*=4.02$ \cite{Delgadillo-Blando:2012skh}. In other words, the region $2/9<\tau_S<1/4$ corresponds to the Yang-Mills matrix phase for all values of $\tilde{\alpha}$.

\item In fact for $\tau_S>2/9$ the geometric fuzzy sphere background is a metastable state with an observable decay to the Yang-Mills matrix background.  This decay is not observable for $\tau_S=2/9$ although the fuzzy sphere is not the true ground state even here.
\item In the Yang-Mills matrix phase the ground state is given by  $\varphi=\varphi_0=0$ and fluctuations are insensitive to the value of  $\tilde{\alpha}$ and are dominated by commuting matrices. In fact, in this phase the matrix model (\ref{action1}) is dominated by the Yang-Mills term \cite{OConnor:2012vwc}.
\item More precisely, the Yang-Mills matrix phase is characterized by a joint eigenvalue distribution, for the three matrices $C_1$, $C_2$ and $C_3$, which is uniform inside a solid ball of some radius $R=2.0$ in $\mathbb{R}^3$. The eigenvalue distribution of a single matrix is then given by the so-called parabolic law, viz \cite{Berenstein:2008eg,Filev:2013pza,OConnor:2012vwc,Filev:2014jxa}
\begin{eqnarray}
\rho(x)=\frac{3}{4R^3}(R^2-x^2).
\end{eqnarray}
\item The transition from the geometric fuzzy sphere phase to the Yang-Mills matrix phase is of an exotic character in the sense that by crossing the coexistence curve at fixed $\tau_S$ from the fuzzy sphere side we encounter divergent specific heat with critical exponent equal $1/2$.  However, by crossing the coexistence curve at fixed $\tilde{\alpha}>\tilde{\alpha}_*=4.02$ we find no critical fluctuations and the transition is associated with a continuous internal energy and discontinuous specific heat.
\end{itemize}
The description of the phase structure of the noncommutative pseudo-sphere $\mathbb{H}^2_{\theta}$ using the effective potential (\ref{effe2}) is formally identical to the fuzzy sphere case. However, here we have at our disposal only the effective potential (\ref{effe2}) since Monte Carlo simulations of the infinite-dimensional Lorentzian  matrix model (\ref{action2}) are very difficult if not impossible. Nevertheless, the phase structure   of the noncommutative pseudo-sphere $\mathbb{H}^2_{\theta}$  can be summarized as follows:
\begin{itemize}

\item The solution $\varphi_H=\varphi_0=0$ (the Yang-Mills or matrix phase) is the global minimum (ground state) of the system in the regime $\tau_H>1/4$. The solution $\varphi=\varphi_-\neq 0$  (the geometric or noncommutative pseudo-sphere phase) is the global minimum in the regime $0<\tau_H<1/4$. The model has no ground state for $\tau_H<0$, i.e. $\beta_H>0$. The two global minima $D_a=0$ and $D_a=\phi_-K_a$ are separated by a potential barrier whose maximum height is reached at the local maximum $\varphi_+$.


\item In the regime of the noncommutative pseudo-sphere the expansion of the matrix model around the noncommutative background $D_a=\phi_-K_a$ gives a noncommutative gauge theory which also includes coupling to a normal scalar field. 

\item 
In the regime $2/9<\tau_H<1/4$  the noncommutative pseudo-sphere becomes unstable. The coexistence curve between the geometric noncommutative pseudo-sphere phase and the Yang-Mills matrix phase asymptotes therefore to the line $\tau_H=2/9$.
\item If we include the logarithmic quantum correction the potential becomes unbounded from below near $\varphi_H=\varphi_0=0$, i.e. the effective potential   (\ref{effe2}) is really valid only in the noncommutative pseudo-sphere phase $\varphi_H=\varphi_-\neq 0$.

\item In the quantum case the minimum $\varphi_-$ (corresponding to the geometric noncommutative pseudo-sphere phase) becomes a function of both $\tau_H$ and $\tilde{\kappa}^4=4\kappa^4 k(k-1)$. The critical coexistence curve exists therefore in the $(\tau_H,\tilde{\kappa})$ plane where the local minimum $\varphi_-$ disappears. This curve is given by equation (\ref{ec}) with the substitution $\tilde{\alpha}\longrightarrow\tilde{\kappa}$, $\tau_S\longrightarrow\tau_H$. 
\item Hence, as the temperature $T_H$ is increased the noncommutative pseudo-sphere phase evaporates to a pure matrix phase with no background geometrical structure, i.e. the geometry condenses or emerges only as the system cools.
\item It is also expected that the coexistence curve between the geometric noncommutative pseudo-sphere phase (low temperatures) and the Yang-Mills matrix phase (high temperatures) for $2/9<\tau_S<1/4$ will asymptote very rapidly to the line $\tau_H=2/9$ for $\tilde{\kappa}>\tilde{\kappa}_*=4.02$. In other words, the region $2/9<\tau_H<1/4$ corresponds to the Yang-Mills matrix phase for all values of $\tilde{\kappa}$.

\item It is also conjectured that fluctuations in the Yang-Mills matrix phase are insensitive to the value of  $\tilde{\kappa}$ and are dominated by commuting matrices, i.e. the matrix model (\ref{action2}) is dominated by the Yang-Mills term. As a consequence the Yang-Mills matrix phase is characterized by a joint eigenvalue distribution which is uniform inside a solid ball of some radius.

\end{itemize}

\section{Phase diagram of  ${\rm AdS}^2_{\theta}\times\mathbb{S}^2_N$}

The determination of the precise content of the phase diagram of the noncommutative near-horizon geometry ${\rm AdS}^2_{\theta}\times\mathbb{S}^2_N$, where the noncommutative anti-de Sitter spacetime ${\rm AdS}^2_{\theta}$ is Wick-rotated into the pseudo-sphere $\mathbb{H}^2_{\theta}$, requires the computation of the effective potential in the background (\ref{HS}).  Monte Carlo simulations are useless in this case since the basic action here given by (\ref{action3}) is an infinite-dimensional (from the perspective of the symmetry group $SO(1,2)$) and Lorentzian (from the perspective of the embedding spacetime) Yang-Mills matrix model. In some sense the Yang-Mills matrix model (\ref{action3}) is genuinely Euclidean only in the four-dimensional geometric phase.

Let us simply start by writing the classical potential computed using the action (\ref{action3}) in the background configuration (\ref{HS}). The sphere and pseudo-sphere sectors are not geometrically entangled at the classical level and thus we obtain
\begin{eqnarray}
\frac{V_{HS}}{2N_{\rm S}^2N_{\rm H}^2}
&=&\bar{\alpha^4}\bigg[\frac{1}{4}\varphi_S^4-\frac{1}{3}\varphi_S^3+\frac{1}{2}\tau_S\varphi_S^2\bigg]\nonumber\\
&+&\bar{\kappa^4}\bigg[\frac{1}{4}\varphi_H^4-\frac{1}{3}\varphi_H^3+\frac{1}{2}\tau_H\varphi_H^2\bigg].\label{classi}
\end{eqnarray}
The order parameters are still given by the two scalar fields $\varphi_S$ and $\varphi_H$ which are measuring the sizes of the sphere and pseudo-sphere respectively. Also recall that the scaling of the deformation parameters (gauge coupling constants)  in four dimension are given by $\bar{\alpha}^4=\tilde{\alpha}^4/4N=\alpha^4N/4$ and  $\bar{\kappa}^4=\tilde{\kappa}^4/4N=\kappa^4N/4$.

Before we sketch the calculation of the quantum effective potential we can immediately state the possible phases of  the noncommutative near-horizon geometry  ${\rm AdS}^2_{\theta}\times\mathbb{S}^2_N$ in the space $(\tau_S,\tau_H,\bar{\alpha},\bar{\kappa})$. The expected phases in this case are as follows:
\begin{itemize}
\item A Yang-Mills matrix phase with no background geometrical structure which is expected at high temperature (both $\bar{\alpha}$ and $\bar{\kappa}$ approach zero) or $2/9<\tau_{S,H}<1/4$.
\item A $2-$dimensional geometric fuzzy sphere phase ($\bar{\alpha}$  approaches infinity, $\bar{\kappa}$ approaches zero and  $0<\tau_{S,H}<2/9$).
\item A $2-$dimensional geometric noncommutative pseudo-sphere phase ($\bar{\alpha}$  approaches zero, $\bar{\kappa}$ approaches infinity and  $0<\tau_{S,H}<2/9$).
\item A $4-$dimensional geometric noncommutative near-horizon geometry  ${\rm AdS}^2_{\theta}\times\mathbb{S}^2_N$ phase at low temperature (both $\bar{\alpha}$ and $\bar{\kappa}$ approach infinity and  $0<\tau_{S,H}<2/9$).
\end{itemize}
However, in order to have a single unified temperature $T_{\rm HS}$ we must have a single unified gauge coupling constant $g_{\rm HS}$ as in equation (\ref{gHS}). This together with the regularization (\ref{reg}) and (\ref{Hreg}) allows us to set  
\begin{eqnarray}
\alpha=\kappa.
\end{eqnarray}
The temperature is then given by $T_{\rm HS}\equiv 1/\bar{\alpha}^4=1/\bar{\kappa}^4=g_{\rm HS}^2$ and the phase diagram becomes three-dimensional in the space $(\tau_S,\tau_H,\bar{\alpha}=\bar{\kappa})$. 

The calculation of the effective potential on the noncommutative near-horizon geometry  ${\rm AdS}^2_{\theta}\times\mathbb{S}^2_N$ which is based on the matrix model (\ref{action3}) is much more involved than the analogous calculation on the fuzzy   $\mathbb{S}^2_N\times\mathbb{S}^2_N$ which is done in \cite{Castro-Villarreal:2005pes,Ydri:2016osu}. The difficulty can be traced to the fact that the sphere and the pseudo-sphere sectors become entangled quantum-mechanically through the third term in the action (\ref{action3}). This geometric quantum entanglement is of course essential for the emergence of a four-dimensional space.  However, there exists a special case where this geometric quantum entanglement can be removed while keeping the background emergent geometry four-dimensional. A very important situation is the case when the coupling constants $\tau_S=\beta_S/\alpha^2$ and $\tau_H=-\beta_H/\kappa^2$  are identical, viz 
\begin{eqnarray}
\tau_S=\tau_H\iff \beta_S=-\beta_H.\label{beta}
\end{eqnarray}
From the actions (\ref{action1}), (\ref{action2}) and (\ref{action3}) it is obvious that these coupling constants couple to the radii of the fuzzy sphere $\mathbb{S}^2_N$ and the noncommutative pseudo-sphere $\mathbb{H}^2_{\theta}$ given respectively by ${\rm Tr}_SC_a^2$ and ${\rm Tr}_HD_a^2$.

In the special case (\ref{beta}) the $2-$dimensional geometric phases are now expected to disappear and we end up with a single phase transition between a geometric noncommutative near-horizon geometry  ${\rm AdS}^2_{\theta}\times\mathbb{S}^2_N$ phase and a Yang-Mills matrix phase.

Let now $A_{\alpha}$ stands for the sphere and the pseudo-sphere configurations, i.e. $A_{\alpha}=(A_a^S,A_a^H)$ where $A_a^S\equiv C_a$ and $A_a^H\equiv D_a$. In the Feynman-'t Hooft background field method we decompose the field as $A_{\alpha}=B_{\alpha}+Q_{\alpha}$ where $Q_{\alpha}$ stands for the sphere and the pseudo-sphere quantum fluctuations and $B_{\alpha}$ stands for the sphere and the pseudo-sphere background fields which solve the classical equations of motion, i.e. $B_{\alpha}=(B_a^S,B_a^H)$ where $B_a^S\equiv C_a=\varphi_S\hat{x}_a$ and $B_a^H\equiv D_a=\varphi_H\hat{X}_a$. We expand the classical action (\ref{action3}) around the background fields $B_{\alpha}$ and keep only terms up to quadratic in the fluctuation fields $B_{\alpha}$. In other words, we write the action (\ref{action3}) in the Gaussian form 
\begin{eqnarray}
S_{\rm HS}[A]&=&S_{\rm HS}[B]+N_{\rm H}N_{\rm S}{\rm Tr}_{\rm H}{\rm Tr}_{\rm S}Q_{\alpha}\tilde{\Omega}_{\alpha\beta}Q_{\beta}+O(Q^3).\label{quadratic}\nonumber\\
\end{eqnarray}
The linear term vanishes by the classical equations of motion. The local symmetry of this action (which is $U(N)$ by the regularization of anti-de Sitter spacetime employed here in this article)  is gauge-fixed using the Lorentz gauge  
\begin{eqnarray}
[B_{\alpha},Q_{\alpha}]\sim [B_a^S,Q_a^S]+\eta^{ab}[B_a^H,Q_b^H]=0.
\end{eqnarray}
We thus add to the above quadratic action (\ref{quadratic}) the usual gauge fixing and Faddeev-Popov terms given by
\begin{eqnarray}
S&\sim& -N_{\rm H}N_{\rm S}{\rm Tr}_{\rm H}{\rm Tr}_{\rm S}\frac{[B_{\alpha},Q_{\alpha}]^2}{2\xi}\nonumber\\
&+&N_{\rm H}N_{\rm S}{\rm Tr}_{\rm H}{\rm Tr}_{\rm S}c[B_{\alpha},[B_{\alpha},b]].
\end{eqnarray}
We will employ the Feynman gauge $\xi=1$. We also note that the gauge-covariant Laplacian operator ${\cal B}^2$ is the sum of the sphere and the pseudo-sphere Laplacian operators, viz
\begin{eqnarray}
{\cal B}^2(f)=[B_{\alpha},[B_{\alpha},f]]\sim [B_a^S,[B_a^S,f]]+\eta^{ab}[B_a^H,[B_b^H,f]].\nonumber\\
\end{eqnarray}
By performing the Gaussian path integral we obtain the one-loop effective action
\begin{eqnarray}
\Gamma_{\rm HS}[B]&=&S_{\rm HS}[B]+\frac{1}{2}Tr\log \Omega-Tr\log{\cal B}^2.
\end{eqnarray}
The first term gives the classical potential (\ref{classi}). The most important term in the full gauge-covariant Laplacian operator $\Omega$ is of the frm 
\begin{eqnarray}
\Omega_{\alpha\beta}={\cal B}^2\delta_{\alpha\beta}+....\label{om}
\end{eqnarray}
As it turns out, all the other terms in the gauge-covariant Laplacian operator $\Omega$, when evaluated in the configuration (\ref{HS}), i.e. in    $B_a^S=\varphi_S\hat{x}_a$ and $B_a^H=\varphi_H\hat{X}_a$, and using the Feynman gauge $\xi=1$, are diagonal in the total Hilbert space associated with the tensor product of the fuzzy sphere and the noncommutative pseudo-sphere Hilbert spaces. These terms are also subleading compared to the first term written in (\ref{om}). The geometric quantum entanglement between the sphere and the pseudo-sphere sectors is then only found in the gauge-covariant Laplacian operator ${\cal B}^2$. Indeed, we compute in the configuration  (\ref{HS}) the trace
\begin{eqnarray}
Tr\log{\cal B}^2=Tr\log\big(\varphi_S^2\Delta_S+\varphi_H^2\Delta_H^2\big).
\end{eqnarray}
$\Delta_S$ and $\Delta_H$ are essentially the Laplacian operators on the sphere and the pseudo-sphere respectively, namely $\Delta_S=[\hat{x}_a,\hat{x}_a,.]]$ and $\Delta_H=[\hat{X}_a,[\hat{X}^a,]]$.

However, if we assume that the coupling constants $\tau_S$ and $\tau_H$ are equal as in (\ref{beta}) then one can check that the order parameters $\varphi_S$ and $\varphi_H$ must solve identical equations of motion. The choice (\ref{beta}) is also motivated by the form of the metric (\ref{met}). A simplified model which captures the dynamics of the theory is then obtained by simply setting the two order parameters  $\varphi_S$ and $\varphi_H$ equal from the outset, i.e.
\begin{eqnarray}
\varphi_S=\varphi_H\equiv\varphi.
\end{eqnarray}
Thus, we get the logarithmic potential 
\begin{eqnarray}
\frac{1}{2}Tr_dTr\log\varphi^2-Tr\varphi^2=\frac{d}{2}N^2\log\varphi^2-N^2\log\varphi^2.
\end{eqnarray}
Clearly, $d=6$ here. The  ${\rm AdS}^2_{\theta}\times\mathbb{S}^2_N$ effective potential is then given by
\begin{eqnarray}
\frac{V_{HS}}{2N_{\rm S}^2N_{\rm H}^2}
&=&2\bar{\alpha^4}\bigg[\frac{1}{4}\varphi_S^4-\frac{1}{3}\varphi_S^3+\frac{1}{2}\tau_S\varphi_S^2\bigg]+\log\varphi^2.\label{quantum}\nonumber\\
\end{eqnarray}
This effective potential is of the same form as the sphere effective potential (\ref{effe1}) with the substitution
\begin{eqnarray}
\tilde{\alpha}^4\longrightarrow 2\bar{\alpha^4}.\label{sub}
\end{eqnarray}
Hence the discussion of the resulting phase structure proceeds along the same lines. The main result is the fact that as the system cools down the  noncommutative geometry of  ${\rm AdS}^2_{\theta}\times\mathbb{S}^2_N$ emerges. Equivalently, as the temperature $T_{\rm HS}=1/\bar{\alpha}^4=g_{\rm HS}^2$ is increased the geometric noncommutative  ${\rm AdS}^2_{\theta}\times\mathbb{S}^2_N$  phase evaporates to a pure Yang-Mills matrix phase with no background geometrical structure. The coexistence curve is still given by (\ref{ec}) with the substitution (\ref{sub}).

\section{Conclusion}
The quantum gravitational fluctuations about the noncommutative geometry of the fuzzy sphere $\mathbb{S}^2$, the noncommutative pseudo-sphere $\mathbb{H}^2$ and the noncommutative near-horizon geometry ${\rm AdS}^2_{\theta}\times\mathbb{S}^2_N$ are given by IKKT-type Yang-Mills matrix models with additional cubic Myers terms which are given explicitly by the actions (\ref{action1}), (\ref{action2}) and (\ref{action3}).

These matrix models exhibit emergent geometry transitions from a geometric phase (the sphere, the pseudo-sphere and the near-horizon geometry of a Reissner-Nordstrom black hole) to a pure Yang-Mills matrix phase with no background geometrical structure.

This fundamental result is confirmed in the case of the sphere by Monte Carlo simulations of the Euclidean Yang-Mills matrix model (\ref{action1}). But in the other cases we have only at our disposal the effective potential. Indeed, the Monte Carlo method can not be applied to the matrix models (\ref{action2}) and (\ref{action3}) in their current form for two main reasons. First, the Lorentzian signature of the embedding spacetime  (these noncommutative spaces are in fact branes residing in a larger spacetime). Second, the Hilbert spaces ${\cal H}_k$ corresponding to the noncommutative operator algebras are actually infinite-dimensional (which is required by the underlying $SO(1,2)$ symmetry). This was regularized in this article by means of a simple cutoff given by (\ref{reg}) and (\ref{Hreg}) which simply says that only $2k-1$ states in the Hilbert spaces are included turning therefore the infinite-dimensional operator algebras into finite-dimensional matrix algebras.

However, we are quite confident in the results derived from the effective potentials for additional arguments as outlined with some detail in the case of ${\rm AdS}^2_{\theta}$ in the second appendix.

The geometric phases (noncommutative geometry phases) are always characterized by a discrete spectrum whereas the Yang-Mills phase is always characterized by a continuum spectrum. These emergent transitions are also believed to be exotic in the sense that they seem to be critical only as we approach the coexistence curve from the noncommutative geometry phase side.

We also find a possibility for topology change transitions in which space or time directions grow or decay as the temperature is varied. Indeed,  the noncommutative near-horizon geometry ${\rm AdS}^2_{\theta}\times\mathbb{S}^2_N$ can evaporate only partially to a fuzzy sphere $\mathbb{S}^2_N$ (topology change) or to a noncommutative anti-de Sitter spacetime ${\rm AdS}^2_{\theta}$ (emergence of time).

\section{${\rm AdS}^2$ black holes in dilaton gravity}

In this appendix, we would like to re-derive the near-horizon geometry ${\rm AdS}^2\times\mathbb{S}^2$ from a dilatonic action principle and briefly discuss its relation to dilaton gravity in two dimensions.

We start with dilaton gravity theory in four dimensions given by the following action \cite{Cadoni:1994uf,Cadoni:1993rn,Garfinkle:1990qj,Giddings:1992kn}

\begin{eqnarray}
S=\int d^4x \sqrt{-{\rm det}g^{(4)}} e^{-2\phi}(R^{(4)}-F_{\mu\nu}F^{\mu\nu}).
\end{eqnarray}
The corresponding spherically symmetric non-singular black hole solution is given by 


\begin{eqnarray}
ds^2=-(1-\frac{r_+}{r})dt^2+\frac{dr^2}{(1-\frac{r_+}{r})(1-\frac{r_-}{r})}+r^2d\Omega_2^2.\label{4dBH}
\end{eqnarray}

\begin{eqnarray}
e^{2(\phi-\phi_0)}=\frac{1}{\sqrt{1-\frac{r_-}{r}}}.
\end{eqnarray}
The temperature and the entropy of the black hole are given on the other hand by the relations \cite{Cadoni:1994uf}

\begin{eqnarray}
T=\frac{1}{4\pi r_+}\sqrt{1-\frac{r_-}{r_+}}~,~S=\pi r_+^2.
\end{eqnarray}
The radii $r_{\pm}$ are given explicitly by
\begin{eqnarray}
2M=r_+~,~ Q_M^2=\frac{3}{4}r_+r_-.
\end{eqnarray}
The extremal limit $T\longrightarrow 0$ of this black hole configuration is then given by $r_+=r_-=Q=2Q_M/\sqrt{3}$ or equivalently $M=Q_M/\sqrt{3}$.


For the extremal solution $r_+=r_-=Q$ we introduce the coordinates

\begin{eqnarray}
r=Q(1+\frac{4\lambda^2}{z^2})~,~t=\frac{QT}{\lambda}.
\end{eqnarray}
The metric and the dilaton in the near-horizon limit $\lambda\longrightarrow 0$ take then the form

\begin{eqnarray}
ds^2=\frac{4Q^2}{z^2}(-dT^2+dz^2)+Q^2d\Omega_2^2.
\end{eqnarray}

\begin{eqnarray}
e^{2(\phi-\phi_0)}=\frac{z}{2\lambda}.
\end{eqnarray}
This shows explicitly that the near-horizon geometry of the extremal black hole is indeed ${\rm AdS}^2\times\mathbb{S}^2$.

We can perform a spherical reduction of this solution by decomposing the metric as follows

\begin{eqnarray}
ds^2&=&g_{\mu\nu}^{(4)}dx^{\mu}dx^{\nu}\nonumber\\
&=&g_{ab}^{(2)}dx^adx^b+\Phi^2(x^a)\gamma_{ij}dn^idn^j.
\end{eqnarray}
The scalar field $\Phi$ is a dilaton field due to the spherical reduction. We compute then (see \cite{Grumiller:2001ea} and references therein)

\begin{eqnarray}
&&\sqrt{-{\rm det}g^{(4)}}=\Phi^2 \sqrt{-{\rm det}g^{(2)}}\sqrt{{\rm det}\gamma }\nonumber\\
&&R^{(4)}=R^{(2)}-\frac{2}{\Phi^2}(-1+\partial_a\Phi\partial^a\Phi)-\frac{4}{\Phi}\Delta\Phi.
\end{eqnarray}
And hence

\begin{eqnarray}
\int d^4x \sqrt{-{\rm det}g^{(4)}} R^{(4)}&=&4\pi \int d^2x \sqrt{-{\rm det}g^{(2)}} (\Phi^2 R^{(2)}\nonumber\\
&+&2\partial_a \Phi\partial^a\Phi+2).
 \end{eqnarray}
Hence, the action reduces to

\begin{eqnarray}
S&=&4\pi \int d^2x \sqrt{-{\rm det}g^{(2)}} e^{-2\phi}(\Phi^2 R^{(2)}+2\partial_a \Phi\partial^a\Phi\nonumber\\
&+&2-\Phi^2 F^2).
\end{eqnarray}
For Schwarzschild-like coordinates the dilaton field $\Phi$ is given by $\Phi=r$. However, in the current case the spherical reduction is performed on a sphere of constant radius $r= Q=2Q_M/\sqrt{3}$, i.e. $\Phi=Q$. We get then the action (with $\Lambda=1/2Q$)

\begin{eqnarray}
S&=&4\pi Q^2 \int d^2x \sqrt{-{\rm det}g^{(2)}} e^{-2\phi}(R^{(2)}+2\Lambda^2).
\end{eqnarray}
This is called the Jackiw-Teitelboim action \cite{JT} which is one of the most important dilatonic gravity models in two dimensions.

\section{On the non-unitary finite-dimensional representations of $SO(1,2)$} 
The results derived from the sphere effective potential (\ref{effe1}) are strongly corroborated by Monte Carlo simulations of the finite-dimensional Euclidean matrix model (\ref{action1}). However, the Monte Carlo method is not available for us in the case of the infinite-dimensional Lorentzian matrix model (\ref{action2}) and  thus in the case of the pseudo-sphere we have at our disposal, at least in the current article,  only the effective potential (\ref{effe2}) . In this appendix we would like to discuss the validity of the results derived from the pseudo-sphere effective potential (\ref{effe2}) from another quite distinct perspective.

Let us first recall that the sphere and the pseudo-sphere are the Euclidean surfaces embedded respectively in the Euclidean space $\mathbb{R}^3$ and in Minkowski spacetime $\mathbb{M}^{1,2}$ which are given explicitly by the conditions
\begin{eqnarray}
x_1^2+x_2^2+x_3^2=1~,~\mathbb{R}^3.
\end{eqnarray}
\begin{eqnarray}
-X_1^2+X_2^2+X_3^2=-1~,~\mathbb{M}^{1,2}.\label{beng1}
\end{eqnarray}
However, from the perspective of the pseudo-sphere matrix model (\ref{action2}) the embedding metric can be either Euclidean or Lorentzian on an equal footing.  Hence, there is the possibility of another space which can be produced by the matrix model (\ref{action2}) given by the two-sheeted hyperboloid which is embedded in $\mathbb{R}^3$ by the condition
\begin{eqnarray}
-Y_1^2+Y_2^2+Y_3^2=-1~,~\mathbb{R}^3.\label{beng2}
\end{eqnarray}
The difference between this two-sheeted hyperboloid (\ref{beng2}) and the pseudo-sphere (\ref{beng2}) lies in its curvature, in its symmetry and its analytic continuation into the sphere, viz  \cite{Bengtsson} 
\begin{itemize}
\item The pseudo-sphere $\mathbb{H}^2$ (like anti-de Sitter spacetime ${\rm AdS}^2$) is a space of a constant negative scalar curvature whereas the two-sheeted hyperboloid (\ref{beng2}) is  a space with a non-constant curvature.
\item Furthermore, the two vector fields $J_{i1}=X_i\partial_1+X_1\partial_i$, $i=2,3$ leave the two-sheeted hyperboloid (\ref{beng2}) invariant but they do not leave invariant the Euclidean metric $ds^2=dX_1^2+dX_2^2+dX_3^3$. Thus, these vector fields are not Killing vectors of the induced metric on the surface (\ref{beng2}), i.e. we do not have $SO(1,2)$ symmetry in this case. In contrast, these two vector fields $J_{i1}$ leave the pseudo-sphere  (\ref{beng1}) and the Minkowski metric  $ds^2=-dX_1^2+dX_2^2+dX_3^3$ invariant at the same time and thus they are Killing vectors of the induced metric on the pseudo-sphere, i.e. we have exactly $SO(1,2)$ symmetry in this case.
\item The two-sheeted hyperboloid (\ref{beng2}) is related to the sphere by the double Wick rotation $Y_i\longrightarrow -i Y_i$, $i=2,3$. As it turns out, the Lorentzian pseudo-sphere matrix model (\ref{action2}) goes to the Euclidean sphere matrix model (\ref{action1}) by the same double Wick rotation. 
\end{itemize}
The two-sheeted hyperboloid (\ref{beng2}) space is then seen to lie midway between the sphere and the pseudo-sphere. Indeed, it is observed in this case that $SO(3)$ is the symmetry group of the embedding metric while $SO(1,2)$ is the symmetry group of the embedded surface.

As we have already said, the matrix model (\ref{action2}) does not know a priori about the embedding metric in the sense that we can obtain from it, as a classical background solution, the following noncommutative two-sheeted hyperboloid given by
\begin{eqnarray}
-\hat{Y}_1^2+\hat{Y}_2^2+\hat{Y}_3^2=-1~,~\hat{Y}_a=\kappa K_a.\label{beng3}
\end{eqnarray}
The noncommutative pseudo-sphere $\mathbb{H}^2_{\theta}$ is given by a similar equation, viz
\begin{eqnarray}
-\hat{X}_1^2+\hat{X}_2^2+\hat{X}_3^2=-1~,~\hat{X}_a=\kappa K_a.
\end{eqnarray}
Indeed, the Casimir operator $C=-K_1^2+K_2^2+K_3^2$ can be negative equal $-k(k-1)$ for both the discrete series $D_k^{\pm}$ (the noncommutative pseudo-sphere) and the finite-dimensional  series $F_k$. The finite-dimensional series $F_k$ is non-unitary corresponding to the $su(2)$ spin quantum number $j=k-1$, i.e. the $su(1,1)$ and $su(2)$ representations are exactly  related by the double Wick rotation $K_i\longrightarrow -iK_i$, $i=2,3$.

Thus, the two-sheeted hyperboloid (\ref{beng2}) must be quantized by means of the non-unitary finite-dimensional irreducible representations $F_k$ of $SO(1,2)$, i.e. $K_a$ in (\ref{beng3}) are the generators of $SO(1,2)$ in the non-unitary finite-dimensional irreducible representations $F_k$.

We conclude that symmetry and Lorentzian signature together correspond to unitary representations which act generically on infinite-dimensional Hilbert spaces. The cutoff used for the pseudo-sphere in this article interferes clearly with the underlying symmetry and as such it probably entails also a conflict with unitarity. However, the pseudo-sphere effective potential (\ref{effe2}) seems to be valid as the two-sheeted hyperboloid (\ref{beng2}) provides a controllable link between the sphere and the pseudo-sphere cases in which the interplay between Lorentzian/Euclidean signature, finite-/infinite-dimensional Hilbert spaces and unitary/non-unitary representations is very clear and very intriguing.

\end{document}